\documentstyle[12pt]{article}
\renewcommand{\theequation}{\arabic{equation}}
\newcommand{\EQ}{\begin{equation}}
\newcommand{\sh}{\sinh}

\newcommand{\ch}{\cosh}
\newcommand{\EN}{\end{equation}}

\newcommand{\ket}[1]{\left|#1\right\rangle}      

\newcommand{\bear}{\begin{eqnarray}}
\newcommand{\ear}{\end{eqnarray}}

\begin{document}

\topmargin 0pt
\oddsidemargin 5mm
\newcommand{\NP}[1]{Nucl.\ Phys.\ {\bf #1}}
\newcommand{\PL}[1]{Phys.\ Lett.\ {\bf #1}}
\newcommand{\NC}[1]{Nuovo Cimento {\bf #1}}
\newcommand{\CMP}[1]{Comm.\ Math.\ Phys.\ {\bf #1}}
\newcommand{\PR}[1]{Phys.\ Rev.\ {\bf #1}}
\newcommand{\PRL}[1]{Phys.\ Rev.\ Lett.\ {\bf #1}}
\newcommand{\MPL}[1]{Mod.\ Phys.\ Lett.\ {\bf #1}}
\newcommand{\JETP}[1]{Sov.\ Phys.\ JETP {\bf #1}}
\newcommand{\TMP}[1]{Teor.\ Mat.\ Fiz.\ {\bf #1}}
     
\renewcommand{\thefootnote}{\fnsymbol{footnote}}
     
\newpage
\setcounter{page}{0}
\begin{titlepage}     
\begin{flushright}
UFSCAR-99-20
\end{flushright}
\vspace{0.5cm}
\begin{center}
\large{ Integrable  
supersymmetric correlated electron chain with open
 boundaries } \\
\vspace{1cm}
\vspace{1cm}
 {\large M.J. Martins$^{1}$  and X.-W. Guan$^{1,2}$} \\
\vspace{1cm}
\centerline{\em ${}^{1}$ Departamento de F\'isica, Universidade Federal de S\~ao Carlos}
\centerline{\em Caixa Postal 676, 13565-905, S\~ao Carlos, Brasil}
\centerline{ \em and }
\centerline{\em ${}^{2}$ Department of Physics, Jilin University}
\centerline{\em Changchun 130023, P. R. China}
\vspace{1.2cm}   
\end{center} 
\begin{abstract}
We construct an extended Hubbard model 
with open boundaries from a $R$-matrix
based on the $U_q[Osp(2|2)]$ superalgebra. We study the reflection
equation and find two classes of diagonal solutions.
The corresponding
one-dimensional open Hamiltonians are diagonalized by means of the 
Bethe ansatz approach.

\end{abstract}
\vspace{.2cm}
\centerline{PACS numbers: 71.27.+a, 75.10.Jm}
\vspace{.2cm}
\centerline{June 1999}
\end{titlepage}

\renewcommand{\thefootnote}{\arabic{footnote}}
\section{Introduction}
In recent years there has been substantial research devoted to the study of
(1+1)-dimensional solvable lattice models with integrable boundary conditions.
In one-dimensional theory with factorized scattering, the
boundary effects can be described by scattering
matrices satisfying the so-called reflection equation \cite{CR}.
A systematic approach to construct such models has been developed
by Sklyanin \cite{SK} who has investigated the six vertex model 
with boundary fields. Subsequently,
this scheme has been generalized to handle  a rather
general class of models based on
Lie algebras \cite{NE,DE} (see also ref. \cite{MV}). By now, variants of this
method have been extensively used in the literature to study various 
integrable quantum field theories and lattice statistical 
mechanics models with open  boundaries. See refs.\cite{QF,SM} just to mention
few examples.

Of particular interest are supersymmetric 
generalizations of the Hubbard model
\cite{EKS,UMO,MA}  due to their possible relevance in 
describing strongly correlated
electron systems.  These models are often derived from supersymmetric
solutions of the Yang-Baxter equation invariant by the $gl(2|1)$ and $Osp(2|2)$
Lie superalgebras \cite{DEG,AU}. A successful example is the supersymmetric
free-parameter model with open boundaries constructed from the $R$-matrix
associated to the four dimensional representation of $gl(2|1)$ \cite{AU1}.
The purpose of this paper is to perform similar
task for another $Osp(2|2)$ $R$-matrix solution found some time ago by Deguchi
et al \cite{DEG}. This latter model, however, appears to 
be specially interesting, since its Hamiltonian provide a lattice
regularization of a relevant 
integrable double sine-Gordon model \cite{SA}.
This continuum field theory with open boundary is known to
describe  tunneling effects 
in quantum wires \cite{SA1}. Therefore, the
Bethe ansatz results for open boundaries we shall derive here could be a
useful non-perturbative tool to investigate this condensed matter
system as well.

This paper is organized as follows. The next section is concerned with diagonal
solutions of the reflection equations  \cite{CR,SK} associated with a particular
$Osp(2|2)$ $R$-matrix. We found two families of one parameter solutions leading to
four possible choices of boundary conditions for integrable open chain Hamiltonians.
In section 3 we formulate their Bethe ansatz solutions in terms of 
the coordinate
Bethe ansatz approach. Section 4 is reserved 
for our conclusions and final remarks. For completeness, 
in appendix A we present another possible $R$-matrix embedding and discuss its
boundary behaviour.
     
\section{The $Osp(2|2)$ $R$-matrix and related $K$-matrices}

The ``bulk'' part of (1+1)-dimensional integrable system with a boundary
is governed by the two-particle scattering  matrix
$R(\lambda)$ 
satisfying the Yang-Baxter equation. The spectral parameter
$\lambda$ plays the role of the difference of the particles pseudomomenta.
The boundary effects 
are described in terms of 
two boundary $K_{\pm}(\lambda)$ matrices that characterize
the scattering interactions at the boundary \cite{CR,SK}. 
The compatibility of the reflections and particle scatterings leads us
to an algebraic condition, the so-called
reflection equation \cite{CR,SK}
\footnote{ We are assuming the ordinary non-graded boundary framework, since
we shall look only for diagonal solutions. For an application of the graded
formalism to non-diagonal solutions, see for instance ref. \cite{AU2}.}
\begin{eqnarray}
R_{12}(\lambda-\mu ) \stackrel{1}{K}(\lambda ) R_{21}(\lambda+\mu)
 \stackrel{2}{K}(\mu ) =  
 \stackrel{2}{K}(\mu )
 R_{12}(\lambda +\mu ) \stackrel{1}{K}(\lambda ) R_{21}(\lambda -\mu)
\label{REF}
\end{eqnarray}

Here we will also require that 
$R$-matrix $R_{12}(\lambda )$
satisfies, besides the standard properties of
regularity and unitarity, certain extra relations denominated
$PT$ and  crossing symmetries, namely
\begin{equation}
PT-symmetry: P_{12}R_{12}(\lambda )P_{12}=R^{t_1t_2}_{12}(\lambda )
\label{PT}
\end{equation}
\begin{equation}
crossing-symmetry: R_{12}(\lambda )=\frac{\rho(\lambda)}{\rho(-\lambda-\eta)}\stackrel{1}
{V}R_{12}^{t_2}(-\lambda -\eta ){\stackrel{1}{V^{-1}}}
\label{CRO}
\end{equation}
where $P_{12}$ is the exchange operator, $t_k$ denotes the transpose in the
space $k$, $\eta $ is the crossing parameter, $V$ is related to the crossing
matrix by $M=V^tV$ and $\rho(\lambda)$ is a convenient normalization function.
 
When these properties are fulfilled one 
can follow the scheme devised by Mezincescu
and Nepomechie \cite{NE}. One of the boundary matrices, 
say the $K_{-}(\lambda)$ matrix,
is obtained by solving the reflection equation (1). The other 
$K_{+}(\lambda)$ matrix is automatically determined by the following isomorphism
\begin{equation}
K_+(\lambda )=K_{-}^{t}(-\lambda -\eta ) M
\end{equation}

Before proceeding, we remark that the original $Osp(2|2)$ $R$-matrix given
in ref. \cite{DEG} is not $PT$-symmetric. However, this difficulty 
is easily circumvent by applying a convenient unitary transformation. It turns
out that the $PT$-symmetric $Osp(2|2)$
$R$-matrix can be presented in the following form
\begin{eqnarray}
R(\lambda ) & = & \sum _{\alpha =1}^{4}\frac{q}{q_{\alpha }}
\frac{e^{\lambda}-q^2_{\alpha}}
{1-e^{\lambda}q^2}E_{\alpha \alpha }
\otimes E_{\alpha \alpha }+\frac{q(e^{\lambda}-1)}{1-e^{\lambda}q^2}
\sum _{\mbox{\scriptsize $\begin{array}{c}\alpha,\beta=1\\
\alpha \neq \beta,\beta^{'}\end{array}$}}^{4}
E_{\alpha \alpha }\otimes E_{\beta \beta } 
\nonumber \\
&& +\frac{1-q^2}{1-e^{\lambda}q^2} 
[e^{\lambda}\sum _{\mbox{\scriptsize $\begin{array}{c}\alpha,\beta=1\\
\alpha < \beta,\alpha \neq \beta^{'}\end{array}$}}^{4}+
\sum _{\mbox{\scriptsize $\begin{array}{c}\alpha,\beta=1\\
\alpha > \beta,\alpha \neq \beta^{'}\end{array}$}}^{4} ]
E_{\alpha \beta }\otimes E_{\beta \alpha }+\frac{e^{\lambda}-1}
{1-e^{\lambda}q^2}\sum _{\alpha =1}^{4}a_{\alpha \beta }(\lambda)
E_{\alpha \beta }\otimes E_{\alpha ^{'}\beta ^{'}} \nonumber \\
\end{eqnarray}
where $q$ is the deformation parameter, $E_{\alpha \beta}$
are the elementary $4\times 4$ matrices and we set
$\alpha ^{'}=5-\alpha $. The functions
$a_{\alpha \beta }(\lambda)$ are
\begin{equation} 
a_{\alpha \beta} (\lambda)  =
\left\{\begin{array}{ll}\frac{\mbox{\scriptsize $q$}}{\mbox{\scriptsize $q_{\alpha }$}}
\frac{\mbox{\scriptsize $q^2_{\alpha }e^{\lambda}+1$}}{\mbox{\scriptsize $e^{\lambda}+1$}} &
\alpha =\beta ,\\ e^{\lambda} \left[\mbox{\scriptsize $\varepsilon _{\alpha }\varepsilon _{\beta }$}
\frac{\mbox{\scriptsize $1-q^2$}}{\mbox{\scriptsize $1+e^{\lambda}$}}+
\frac{\mbox{\scriptsize $1-q^2$}}{\mbox{\scriptsize $e^{\lambda}-1$}}\mbox{\scriptsize $\delta _{\alpha ,\beta ^{'}}$}\right] &
 \alpha < \beta\\
-\mbox{\scriptsize $\varepsilon _{\alpha }^{-1}\varepsilon _{\beta }^{-1}$}\frac{\mbox{\scriptsize $1-q^2$}}
{\mbox{\scriptsize $1+e^{\lambda}$}}+
\frac{\mbox{\scriptsize $1-q^2$}}{\mbox{\scriptsize $e^{\lambda}-1$}}\mbox{\scriptsize $\delta _{\alpha ,\beta ^{'}}$} & 
\alpha > \beta \end{array}  \right.
\end{equation}
and the parameters $q_{\alpha }$ and $\varepsilon _{\alpha }$ are defined by
\begin{equation}
q_{\alpha }  = \left\{\begin{array}{ll} q & \alpha =1,4\\ -q^{-1} & \alpha =2,3 \end{array}\right. ,~~~
\varepsilon _{\alpha } = \left\{\begin{array}{ll} 1 & \alpha =1,4\\ -i & \alpha =2,3 \end{array}\right.
\end{equation}

It is not difficult to verify that this $R$-matrix 
satisfies the properties (2,3), where 
\begin{equation}
\eta =  i \pi,~~\rho(\lambda ) = \frac{e^{\lambda}-1}{1-e^{\lambda} q^2},~~~
V  = \left(\matrix{0&0&0& i \cr 0&0&1&0\cr 0&1&0&0 \cr -i &0&0&0\cr }\right) 
\end{equation}

Now we have the basic ingredients to built up integrable boundary models associated
with the $R$-matrix (5). In this paper we are interested to look for diagonal
solutions of the reflection equation, which will play the role of boundary fields in the
context of spin chains. Substituting the ansatz
\begin{equation}
K_-^{(l)}(\lambda,\xi _-)=\left(\matrix{A^{(l)}(\lambda ,\xi _-)&0&0&0\cr
0&B^{(l)}(\lambda ,\xi _-)&0&0\cr 0&0&C^{(l)}(\lambda ,\xi _-)&0 \cr
0&0&0&D^{(l)}(\lambda ,\xi _-)\cr }\right)
\end{equation}
in equation (1), and after some algebra we find two 
classes of diagonal solutions
given by
\begin{eqnarray}
A^{(1)}(\lambda ,\xi _-) & = &
(e^{\lambda }+e^{\xi _-})(e^{\lambda }-e^{\xi _-}q^{-2})
\nonumber \\
B^{(1)}(\lambda ,\xi _-) & = &
(e^{-\lambda }+e^{\xi _-})(e^{\lambda }-e^{\xi _-}q^{-2}) 
 \nonumber \\
C^{(1)}(\lambda ,\xi _-) & = &
(e^{-\lambda }+e^{\xi _-})(e^{\lambda }-e^{\xi _-}q^{-2})
 \nonumber \\
D^{(1)}(\lambda ,\xi _-) & = &
(e^{-\lambda }+e^{\xi _-})(e^{-\lambda }-e^{\xi _-}q^{-2})
\end{eqnarray}
and
\begin{eqnarray}
A^{(2)}(\lambda ,\xi _-) & = &
(e^{\lambda }+e^{\xi _-})(e^{\lambda }-e^{\xi _-}q^{-2})(e^{\lambda }-e^{-\xi _-}q^2)\nonumber \\
B^{(2)}(\lambda ,\xi _-) & = &
(e^{-\lambda }+e^{\xi _-})(e^{\lambda }-e^{\xi _-}q^{-2})(e^{\lambda }-e^{-\xi _-}q^2) \nonumber\\ 
C^{(2)}(\lambda ,\xi _-) & = &
(e^{\lambda }+e^{\xi _-})(e^{\lambda }-e^{\xi _-}q^{-2})(e^{-\lambda }-e^{-\xi _-}q^2) \nonumber \\
D^{(2)}(\lambda ,\xi _-) & = &
(e^{\lambda }+e^{\xi _-})(e^{-\lambda }-e^{\xi _-}q^{-2})(e^{-\lambda }-e^{-\xi _-}q^2)  
\end{eqnarray}

The corresponding $K_+^{(l)}(\lambda ,\xi _+)$ 
matrices are easily derived from the isomorphism (4) with the help of 
crossing property (8). However, it is convenient to introduce suitable 
$K_+^{(l)}(\lambda ,\xi _+)$ matrices to make the interactions at the
boundaries be as much as possible symmetric.  With this in mind, we 
choose the following
expression for these matrices
\begin{equation}
K_{+}^{(l)}(\lambda,\xi_{+})= K_{-}^{(l)}(-\lambda-i\pi, i\pi+2\log[q]-\xi_{+})^{t}M
\end{equation}
where $\xi _{\pm }$ are two independent 
parameters that characterize the
interactions at the right and left ends of the open chain, respectively.

Equipped with the refection matrices $K_{\pm}^{(l)}(\lambda ,\xi_{\pm})$,
an integrable model with open boundary condition
can be obtained through the double-row transfer matrix
$t(\lambda )$ formulated by Sklyanin \cite{SK}
\begin{equation}
t^{(l,m)}(\lambda) ={\rm Tr}_a\left[\stackrel{a}{K}_+^{(m)}(\lambda )
T(\lambda )\stackrel{a}{K}^{(l)}_-(\lambda )T^{-1}(-\lambda )\right],~~ l=m=1,2
\end{equation}
where $T(\lambda )=R_{aL}(\lambda) \cdots R_{a1}$ 
is the standard monodromy matrix of the 
corresponding closed chain with $L$ sites.
We note that the fact $K_{+}^{(l)}(\lambda ,\xi _+)$ can be taken from either
$K_-^{(1)}(\lambda ,\xi _-)$ or $K_-^{(2)}(\lambda ,\xi _-)$ 
gives us four possible
choices of boundary transfer matrices.  In next section we will present the Bethe 
ansatz solution of the corresponding four one-dimensional Hamiltonians with open
boundaries.

\section{Integrable open boundary Hamiltonians}

In order to obtain solvable Hamiltonians 
with open boundaries we have to expand
the double-row transfer matrix $t^{(l,m)}(\lambda )$ up to the second order 
in the spectral parameter $\lambda $ \cite{LI}. This 
is because the relevant term in the 
first order expansion, which is proportional
to the trace of $K_+^{(l)}(0)$, turns out to be
zero  
for both classes of $K$-matrices of section 2.
Following ref. \cite{AU1}, it can be shown that the corresponding Hamiltonian
commuting with $t^{(l,m)}(\lambda )$ is given by
\begin{eqnarray}
 H^{(l,m)} & = & 
 \sum _{j=1}^{L-1}H_{j,j+1}+\frac{\zeta}{2}{
\frac{d}{d \lambda}\stackrel{a}{K_-^{(l)}}}(0)
 +\frac{1}{\varrho^{(m)}}\left\{{\rm Tr}_a\left[{
\frac{d}{d \lambda} \stackrel{a}{K_+^{(m)}}(0) H_{La}}\right]
 \right. \nonumber \\
& &
 \left. \frac{1}{2}{\rm Tr}_a\left[
\stackrel{a}{K_+^{(m)}}(0) \frac{d^2}{d^2 \lambda}R_{aL}(0)P_{aL}\right]+
\frac{1}{2\zeta}{\rm Tr}_a\left[\stackrel{a}{K_+^{(m)}}(0)H_{La}^2\right]\right\}
\end{eqnarray}
where
\begin{equation}
H_{j,j+1}  = 
P_{j,j+1} \frac{d}{d \lambda} R_{j,j+1}(0),~~ R_{12}(0)=  \zeta P_{12},~~
\varrho^{(m)} =
{\rm Tr}_a\left[\frac{d}{d \lambda}{\stackrel{a}{K_+^{(m)}}}(0)\right]
+\frac{2}{\zeta}{\rm Tr}_a\left[\stackrel{a}{K_+^{(m)}}(0)
H_{La}\right]
\end{equation}

We recall that in the derivation of expression (14) it is implicitly 
required that $K_{-}^{(l)}(\lambda, \xi_{-})$ has been normalized by a scalar
function such that $K_{-}^{(l)}(0,\xi_{-})=Id$. 
For both cases (10,11), the only non-trivial 
contributions to the boundaries come
from the first two terms of equation (14). The remaining 
ones are proportional to the identity,
playing the role of irrelevant additive constants. Considering 
the physical motivations of the 
introduction, one would like to express these 
open Hamiltonians in terms of fermionic creation
and annihilation fields 
$c_{j,\sigma}^{\dagger}$ and $c_{j,\sigma}$ acting on the site $j$ and
carrying spin index $\sigma =\pm $. After some algebra, the 
final expression for the Hamiltonians can be written as follows
\begin{equation}
 H^{(l,m)} =
 \sum _{j=1}^{L-1}H_{j,j+1}+U(\xi_{-}) n_{1+}n_{1-}+
 U(\xi_+) n_{L+}n_{L-} 
 +\sum _{\sigma =\pm }{\mu}_{1\sigma }^{(l)}n_{1\sigma}
 +{\mu}_{L\sigma }^{(m)}n_{L\sigma}
\end{equation}
where $n_{j,\sigma }=c_{j,\sigma }^{\dagger } c_{j,\sigma } $ is the number
operator for electrons with spin $\sigma $ on site $j$. The
expression for the
bulk part is  
\begin{eqnarray}
H_{j,j+1} & = &
\sum_{\sigma=\pm} \left [ c_{j, \sigma}^{\dagger} 
c_{j+1, \sigma} +
h.c  \right ] \left [ 1- n_{j, -\sigma}(1+\sigma V_1)
-n_{j+1, -\sigma}(1-\sigma V_1) \right ] 
\nonumber \\
& &
+ V_2 \left[ c_{j,+}^{\dagger} c_{j,-}^{\dagger} 
c_{j+1,-} c_{j+1,+}
- c_{j,+}^{\dagger} c_{j+1,-}^{\dagger} c_{j+1,+} c_{j,-} +h.c. \right] 
+i V_1 \sum_{\sigma=\pm} [n_{i\sigma}-n_{i+1 \sigma}]
\nonumber\\
& &
+ V_2\left [ n_{j,+} n_{j,-} +n_{j+1,+} n_{j+1,-} + n_{j,+} n_{j+1,-} 
+n_{j,-}n_{j+1,+} -\sum_{\sigma=\pm}(n_{j\sigma} +n_{j+1\sigma}) \right ]
\end{eqnarray}
where the couplings $V_1$ and $V_2$ are determined in terms of the parameter
$q=\exp(i\gamma)$ by
\begin{equation}
V_1=  \sin(\gamma),~~ V_2 = \cos(\gamma).
\end{equation}

Turning now to the boundary interactions we found  that 
the on-site Coulomb coupling 
$U(\xi_{\pm})$ is given by
\begin{equation}
U(\xi_{\pm})= -i \frac{\sin(2 \gamma)}{\sinh(\xi_{\pm}/2-i\gamma) \cosh(\xi_{\pm}/2)} 
\end{equation}
while the boundary chemical potentials are
\begin{equation}
\mu_{1\sigma }^{(l)} = 
\left\{\begin{array}{l}
\mu_{1+}^{(1)}=\mu_{1-}^{(1)}=iV_1\frac{e^{-\xi_{-}/2+i\gamma}}{\sinh(\xi_{-}/2-i \gamma)} \\  
 \mu_{1+}^{(2)}=i V_1 \frac{e^{-\xi_{-}/2+i \gamma}}{\sinh(\xi_{-}/2-i \gamma)},~
 \mu_{1-}^{(2)}=-i V_1 \frac{e^{\xi_{-}/2}}{\cosh(\xi_{-}/2)} 
\end{array}\right. 
 \end{equation}
and
\begin{equation}
\mu_{L\sigma }^{(m)} = 
\left\{\begin{array}{ll}\mu_{L+}^{(1)}=
\mu_{L-}^{(1)}=i V_1\frac{e^{\xi_{+}/2-i\gamma}}
{\sinh(\xi_{+}/2-i\gamma)}  \\  
 \mu_{L+}^{(2)}=i V_1 \frac{e^{\xi_{+}/2-i\gamma}}
{\sinh(\xi_{+}/2-i\gamma)},~
 \mu_{L-}^{(2)}=i V_1 \frac{e^{-\xi_{+}/2}}{\cosh(\xi_{+}/2)} 
\end{array}\right. 
 \end{equation}

Our next task is to diagonalize 
the Hamiltonian (16) by the coordinate Bethe ansatz formalism. 
The number of electrons $N_e^{\sigma}$ with spin $\sigma$ are
conserved quantities and they label the 
possible disjoint sectors of the Hilbert space. 
For a sector of a given number 
of particles $N_e=N_e^{+}+N_{e}^{-}$ the Bethe wave function assumes the
following form
\begin{equation}
\ket{\Psi}=\sum _{\mbox{\scriptsize $x_{Q_j},\sigma _j$}} \sum _{P} \mbox{{\rm sgn}}(P)\prod _{j=1}^{N_e}
e^{\mbox{\scriptsize $[ik_{p_j}x_{Q_j}]$}}
A(k_{PQ_1},\cdots, k_{PQ_{N_e} })_{\mbox{\scriptsize 
$\sigma_{Q_1}, \cdots, \sigma_{Q_{N_e}}$ }} 
c^{\dagger}_{\mbox{\scriptsize $ x_{Q_1} $}}\cdots 
c^{\dagger }_{\mbox{\scriptsize $ x_{Q_{N_e}}$}} \ket{0}
\end{equation}
where $\ket{0}$ denotes a reference state containing none particles, 
$1\leq x_{Q_1}\leq x_{Q_2}\leq \cdots \leq x_{\mbox{\scriptsize $
 Q_{N_e}$}}\leq L$ indicate the positions of the electrons,  
 $P$ is the sum over all the permutations of the momenta ($P_1\cdots P_{N_e}$)
 and the symbol $sgn$ 
denotes the sign of the permutation. For configurations  such that
$|x_{Q_i}-x_{Q_j}| \geq 2$
 the Hamiltonian (16) behaves as a free-theory and the
solution of the eigenvalue problem
 $H^{(l,m)}\ket{\Psi}=E^{(l,m)}(L)\ket{\Psi}$ is 
\begin{equation}
 E^{(l,m)}(L)=\sum ^{N_e}_{j=1}2\cos(k_j) 
\end{equation}
up to some additive constants proportional
to the number of electrons $N_e^{\pm }$. It is 
standard in Bethe ansatz approach that configurations in which the
electrons are nearest neighbors, at the same site or 
even at the boundaries
impose constraints on
the amplitudes 
of the wave function. For previous similar computations 
to other models with boundary, see 
refs.\cite{AL,JA}. We
found that such 
consistency condition on the ``bulk'' 
provide us to the following relation
\begin{equation}
 A_{\cdots \mbox{\scriptsize $ \sigma _j,\sigma _i$}\cdots }(\cdots ,k_j,k_i,\cdots )=
 S_{i,j}(k_i,k_j)A_{\cdots \mbox{\scriptsize $ \sigma _i,\sigma _j$}\cdots }(\cdots ,k_i,k_j,\cdots )
\end{equation}
while the reflection at the left and right ends of the chain gives us
\begin{eqnarray}
A_{\mbox{\scriptsize $ \sigma _i,$}\cdots }(-k_j,\cdots ) & = &
S_{\mbox{{\rm l}}}(k_j,P_{{\rm l}\sigma _i}^{(l)})
A_{\mbox{\scriptsize $ \sigma _i,$}\cdots }(k_j,\cdots )\\
A_{\cdots ,\mbox{\scriptsize $ \sigma _i$}}(\cdots, -k_j) & = &
S_{\mbox{{\rm r}}}(k_j,P_{{\rm r}\sigma _i}^{(m)})
A_{\cdots \mbox{\scriptsize $ \sigma _i,$}}(\cdots,k_j)
\end{eqnarray}

The two-body $S$-matrix $S_{i,j}(k_i,k_j)$ connects the scattering amplitudes
between the states $\{(k_i,\sigma _i);(k_j,\sigma _j)\}$ and 
$\{(k_j,\sigma _j^{'});(k_i,\sigma _i^{'})\}$ and its non-null elements are \cite{MA}
\begin{eqnarray}
S_{++}^{++}(\lambda) & = &
S_{--}^{--}(\lambda)=1\\
S_{+-}^{+-}(\lambda) & = & 
S_{-+}^{-+}(\lambda)=
\frac{\sh(\lambda)}{\sh(\lambda+2i\gamma)}\\
S_{+-}^{-+}(\lambda) & = &
S_{-+}^{+-}(\lambda)= 
\frac{\sh(2i \gamma)}{\sh(\lambda+2i\gamma)}
\end{eqnarray}
where the rapidities
$\lambda _j $($ \lambda=\lambda_1-\lambda_2$) 
are related to the momenta $k_j$ by
\begin{equation}
\exp[ik_j] = \frac{\sh(\lambda_j/2 -i\gamma/2)}{\sh(\lambda_j/2 +i\gamma/2)}
\end{equation}

Finally, the boundary scattering matrices are 
\begin{eqnarray}
S_{\mbox{{\rm l}}}(k_j,P_{{\rm l} \sigma }^{(l)})& = &
\frac{1+[P_{{\rm l} \sigma }^{(l)}-2\cos (\gamma )]e^{ik_j}}
{1+[P_{{\rm l}\sigma }^{(l)}-2\cos (\gamma )]e^{-ik_j}},\\
S_{\mbox{{\rm r}}}(k_j,P_{{\rm r}\sigma }^{(m)}) & = &
\frac{1+[P_{{\rm r}\sigma }^{(m)}-2\cos (\gamma )]e^{-ik_j}}
{1+[P_{{\rm r}\sigma }^{(m)}-2\cos (\gamma )]e^{ik_j}} e^{2(L+1)k_j}
\end{eqnarray}
where
\begin{equation}
P_{{\rm l}\sigma }^{(l)} = 
\left\{\begin{array}{l}P_{{\rm l}+}^{(1)}=P_{{\rm l}-}^{(1)}= \frac{\cosh(\xi_{-}/2 +i\gamma)}{\cosh(\xi_{-}/2)}
\\  
 P_{{\rm l}+}^{(2)}=\frac{\cosh(\xi_{-}/2+i\gamma)}{\cosh(\xi_{-}/2)},~
 P_{{\rm l}-}^{(2)}=\frac{\sinh(\xi_{-}/2-2i\gamma)}
{\sinh(\xi_{-}/2-i\gamma)}
\end{array}\right. 
 \end{equation}
 and
\begin{equation}
P_{{\rm r}\sigma }^{(m)} = 
\left\{\begin{array}{l} P_{{\rm r}+}^{(1)}=P_{{\rm r}-}^{(1)}= \frac{\cosh(\xi_{+}/2 +i\gamma)}{\cosh(\xi_{+}/2)}
\\  
 P_{{\rm r}+}^{(2)}=\frac{\cosh(\xi_{+}/2+i\gamma)}{\cosh(\xi_{+}/2)},~
 P_{{\rm r}-}^{(2)}=\frac{\sinh(\xi_{+}/2-2i\gamma)}
{\sinh(\xi_{+}/2-i\gamma)}
\end{array}\right. 
 \end{equation}

So far we managed to solve the charge degrees of freedom of the system 
but we still have to diagonalize the spin sector associated with the
scattering matrices $S_{i,j}(k_i,k_j),  
S_{\mbox{{\rm l}}}(k_i,P_{{\rm l}\sigma }^{(l)})$ 
and $S_{\mbox{{\rm r}}}(k_i,P_{{\rm r}\sigma }^{(m)})$. These 
amplitudes, however,
are easily related to those of
the six-vertex model and
the spin part
of the problem is reduced to 
the diagonalization of an inhomogeneous 6-vertex model with
open boundaries \cite{JA}. In the course 
of solution one has to introduce a second Bethe ansatz
for the spin  rapidities $\mu_j, j=1,\cdots ,N^+_e$. Since this
problem has been
discussed in many different contexts 
in the literature \cite{AL,SK,JA} we restrict ourself
to present only the final 
Bethe ansatz results. For the four possible boundary case, we
find that the pseudomomenta
${\lambda _j}$ and the spin variables ${\mu _j}$ satisfy the following 
nested Bethe ansatz equations
\begin{equation}
\left [ \frac{\sh(\lambda_j/2-i\gamma/2)}{\sh(\lambda_j/2 +i\gamma/2)} 
\right ]^{2L}F(\lambda _j,\xi _{\pm })  = 
 \prod_{k=1}^{N_{e}^{+}} \frac{ \sh(\lambda_j - \mu_k -i\gamma)}
{ \sh(\lambda_j - \mu_k +i\gamma)}
\frac{ \sh(\lambda_j +\mu_k -i\gamma)}{\sh(\lambda_j + \mu_k +i\gamma)},~ 
j=1, \cdots, N_{e} \\
\end{equation}
\begin{eqnarray}
\prod_{k=1}^{N_{e}} \frac{ \sh(\mu_j -\lambda_k -i \gamma)}
{\sh(\mu_j -\lambda_k + i \gamma)}
\frac{ \sh(\mu_j +\lambda_k -i \gamma)}{\sh(\mu_j +\lambda_k + i \gamma)} & = &
G(\mu _j,\xi _{\pm }) 
\prod_{\mbox{\scriptsize $\begin{array}{c} k=1\\
k \neq j \end{array}$}}^{N_e^{+}}
\frac{\sh(\mu_j -\mu_k- 2i \gamma)}{\sh(\mu_j -\mu_k+ 2i \gamma)}
\frac{\sh(\mu_j +\mu_k- 2i \gamma)}{\sh(\mu_j +\mu_k- 2i \gamma)} \nonumber \\
& &
j=1, \cdots, N_{e}^{+}
\end{eqnarray}
where the boundary factors $F(\lambda _j,\xi _{\pm })$ and $G(\mu_j,\xi _{\pm })$
are given by
\begin{eqnarray}
F(\lambda _j,\xi _{\pm }) & = &
\frac{\cosh (\lambda _j/2+i\gamma /2-\xi _-/2)}{\cosh (\lambda _j/2-i\gamma /2+\xi _-/2)}
\frac{\cosh ( \lambda _j/2+i\gamma /2-\xi _+/2)}{\cosh (\lambda _j/2-i\gamma /2+\xi _+/2)}\\
G(\mu _j,\xi _{\pm }) & = &
\left\{\begin{array}{ll}\frac{\cosh ^2(\mu _j-i\gamma )}{\cosh ^2(\mu _j+i\gamma )} &l=m=1\\
-\frac{\sinh (\mu_j-i2\gamma +\xi_+)}{\sinh (\mu_j+i2\gamma -\xi_+)}
\frac{\cosh (\mu_j-i\gamma )}{\cosh (\mu_j+i\gamma )} &l=1,m=2\\
-\frac{\cosh (\mu _j-i\gamma )}{\cosh (\mu _j+i\gamma )}
\frac{\sinh (\mu_j-2i\gamma+\xi_-)}{\sinh (\mu_j-\xi_-+2i\gamma)} & l=2,m=1\\
\frac{\sinh ( \mu _j-i2\gamma +\xi _-)}{\sinh (\mu _j+i2\gamma -\xi _-)}
\frac{\sinh (\mu_j-2i\gamma+\xi_+)}{\sinh (\mu_j+2i\gamma-\xi_+)} & l=m=2 \end{array}\right.
\end{eqnarray}
and, in terms of the rapidities $\lambda_j$, the eigenvalues 
$E^{(l,m)}(L)$ are given (modulo additive constants)
by
\EQ
E^{(l,m)}(L) =  \sum_{i=1}^{N_e} \frac{2 \sin^2(\gamma)}{\cos(\gamma) -\ch(\lambda_i)}
\EN

We close this section commenting on two special open boundary conditions. The
quantum group is obtained from the results for $l=m=2$ in the limits
$ \xi_{-} \rightarrow -\infty $ and $\xi_{+} \rightarrow +\infty$. 
We remark that there is a more transparent way to derive 
such boundary condition, however we have to use 
a different $R$-matrix embedding. For further details 
see Appendix $A$.
Other interesting boundary, concerning critical behaviour, is the 
free-boundary condition. We note that this case is achieved by setting
$\xi_{+}=\xi_{-}=i \pi +2i\gamma$ in the model $l=m=1$.

\section{Conclusions}

We have completed the analysis of the integrability of an interesting 
supersymmetric Hubbard-like model 
in the presence of boundary fields. This was
accomplished by first deriving diagonal solutions of the reflection equation
associated with a particular $U_q[Osp(2|2)$ invariant $R$-matrix. This 
leads us to four boundaries conditions for the corresponding one-dimensional
Hamiltonian, which have been diagonalized by the Bethe ansatz approach.
Quantum-group invariant solutions have been discussed either as a special
limit of the free-parameter $\xi_{\pm}$ or by the analysis of other
possible $R$-matrix embedding.

The Bethe ansatz equations of section 3 provide us a tool to compute
the thermodynamic behaviour and the 
finite-size corrections to the spectrum of the system. In principle,
this allows us to determine the scattering of the physical excitations and
the bulk and the boundary critical properties of the underlying field theory.
These computations could be of interest as an alternative way
to rederive the results 
of ref.\cite{SA1} for the integrable double sine-Gordon model. 
This also opens the possibility to obtain extra information concerning the
operator content of this system  which
should  provide further insight
to the problem of tunneling in quantum wires.

Finally, we mention that one possible generalization of this work
is to investigate operator valued solutions of the reflection equation
associated with the $Osp(2|2)$ $R$-matrix (5) \cite{CHU,AU2,FRA}. This 
will leads us  to an electronic system with 
Kondo impurities \cite{CHU,AU2} which hopefully could be the lattice 
analog of an interesting double sine-Gordon model with Kondo impurity.
We plan to investigate these problems in future publications.

\section*{Acknowledgements}
This  
research has been supported by  
$Fapesp$( Funda\c c\~ao de Amparo a Pesquisa do Estado de S.Paulo) 
and  
partially by $CNPq$ (Brazilian research program). 

\centerline{\bf Appendix A : Other $R$-matrix embedding }
\setcounter{equation}{0}
\renewcommand{\theequation}{A.\arabic{equation}}

The purpose of this appendix 
is to discuss an extra $R$-matrix embedding for the $Osp(2|2)$
vertex model. The $R$-matrix has the same structure of equation (5) but with new weights
$a_{\alpha \beta }(\lambda)$ for $\alpha \neq \beta $, namely
\begin{equation} 
a_{\alpha \beta} (\lambda)=\left\{\begin{array}{ll}\frac{\mbox{\scriptsize $q$}}
{\mbox{\scriptsize $q_{\alpha }$}}
\frac{\mbox{\scriptsize $q^2_{\alpha }e^{\lambda}+1$}}{\mbox{\scriptsize $e^{\lambda}+1$}} &
\alpha =\beta \\ 
e^{\lambda} 
\left[\mbox{\scriptsize $\varepsilon _{\alpha }\varepsilon _{\beta }$}
q^{\mbox{\scriptsize $\stackrel{-}{\alpha }-\stackrel{- }{\beta }$}}
\frac{\mbox{\scriptsize $1-q^2$}}{\mbox{\scriptsize $1+e^{\lambda}$}}+
\frac{\mbox{\scriptsize $1-q^2$}}{\mbox{\scriptsize $e^{\lambda}-1$}}
\mbox{\scriptsize $\delta _{\alpha ,\beta ^{'}}$}\right]&
 \alpha < \beta,\\
-\mbox{\scriptsize $\varepsilon _{\alpha }^{-1}\varepsilon _{\beta }^{-1}$}
q^{\mbox{\scriptsize $\stackrel{-}{\alpha }-\stackrel{-}{\beta }$}}
\frac{\mbox{\scriptsize $1-q^2$}}{\mbox{\scriptsize $1+e^{\lambda}$}}+
\frac{\mbox{\scriptsize $1-q^2$}}{\mbox{\scriptsize $e^{\lambda}-1$}}\mbox{\scriptsize $\delta _{\alpha ,\beta ^{'}}$}&
\alpha > \beta \end{array} \right. 
\end{equation}
where $\varepsilon _1=-\varepsilon _4=q,\varepsilon _2=-\varepsilon _3=i$ and 
$\stackrel{- }{\alpha }$ is defined by
\begin{equation}
 \stackrel{-}{\alpha } = \left\{\begin{array}{ll} \alpha-\frac{1}{2} &1 \leq \alpha \leq 2\\ 
 \alpha +\frac{1}{2} & 3 \leq\alpha \leq 4 \end{array} \right.
\end{equation}

This $R$-matrix satisfies the properties (2,3), but now the crossing matrix $V$
is
\begin{equation}
V  =   \left(\matrix{0&0&0&q^{-1}\cr 0&0&iq^{-1}&0\cr 0&-iq&0&0 \cr q&0&0&0\cr }\right) 
\end{equation}

We note that  some of the new Boltzmann 
weights $a_{\alpha \beta }(\lambda)$ have indeed a different 
functional form as compared to those of equation (6). For periodic boundary 
conditions, however, we have checked that 
such differences are not important as long
as Bethe ansatz analysis is concerned. The corresponding 
Bethe ansatz equations of this
``new'' vertex model (or associated quantum spin chain) 
are precisely the same as that
found in ref.\cite{MA}.
The situation for open boundary conditions is, however, not so rich as 
in section 2.  Although we
managed to find two classes of diagonal 
$K$-matrices solutions, none of them possess a 
free-parameter.  The first solution is the standard quantum-group invariant
one
\begin{equation}
 K_-^{(1)}(\lambda )=Id
 \end{equation}
 while the second class is given by
 \begin{equation}
K_-^{(2)}(\lambda )=\left(\matrix{A_1(\lambda )&0&0&0\cr
0&A_2(\lambda )&0&0\cr 0&0&A_2(\lambda )&0 \cr
0&0&0&A_3(\lambda )\cr }\right) 
\end{equation}
where
\begin{eqnarray}
A_1(\lambda ) & = &
(e^{\lambda }+\epsilon iq^3)(e^{\lambda }-\epsilon iq^{-3})\nonumber \\
A_2(\lambda ) & = &
(e^{-\lambda }+\epsilon iq^3)(e^{\lambda }-\epsilon iq^{-3}) \\
A_3(\lambda ) & = &
(e^{-\lambda }+\epsilon iq^3)(e^{-\lambda }-\epsilon iq^{-3}) \nonumber 
\end{eqnarray}
where $\epsilon =\pm 1$. The
Bethe ansatz solution for such open boundaries follows 
closely the steps of section 3. The only difference
is concerned with
the bulk scattering matrix of the spins degree of freedom. Now, this
matrix possesses the quantum-group
invariant form:
\begin{eqnarray}
S_{++}^{++}(\lambda) & = &
S_{--}^{--}(\lambda)=1,\\
S_{+-}^{-+}(\lambda) & = &
S_{-+}^{+-}(\lambda)= 
\frac{\sh(\lambda )}{\sh(\lambda+2i\gamma)}\\
S_{-+}^{-+}(\lambda)& = & 
e^{\lambda }\frac{\sh(2i\gamma )}{\sh(\lambda+2i\gamma)},\\
S_{+-}^{+-}(\lambda) & = &
e^{-\lambda }\frac{\sh(2i \gamma)}{\sh(\lambda+2i\gamma)}.
\end{eqnarray}

Having this information, it is easy to derive  that the
Bethe ansatz equations associated with the first boundary (A.4)
are of quantum-group type, 
i.e. $F(\lambda_j, \xi_{\pm})=G(\mu _j, \xi_{\pm})=1$. 
Similarly,  
the  Bethe ansatz equations for the second 
boundary (A.5)  gives us $F(\lambda_j,\xi_{\pm})= f_{\epsilon_{-}}(\lambda_j) 
f_{\epsilon_{+}}(\lambda_j) $ and 
$G(\mu _j,\xi_{\pm})=1$ where
\begin{equation}
f_{\epsilon}(\lambda) = 
\left\{\begin{array}{ll} \frac{\cosh[\lambda/2 -i(\gamma+\pi/4)]}{\cosh[\lambda/2+i(\gamma+\pi/4)]}
& \epsilon=+1\\  
-\frac{\sinh[\lambda/2 -i(\gamma+\pi/4)]}{\sinh[\lambda/2+i(\gamma+\pi/4)]}
& \epsilon=-1  
\end{array}\right. 
 \end{equation}

We remark that (A.11)  can be recovered from our previous results
for the ``mixed'' boundary conditions $l=1,m=2$ and $l=2,m=1$ via
fine tuning of the parameters $\xi_{\pm}$. 
The conclusions of this appendix suggest 
that such different embedding may be formulated as
a twisting of Deguchi et al \cite{DEG} original solution
\footnote{We thank J. Links for pointing out this possibility.}.
It  would be
interesting to explore this possibility further since this may 
lead us to
new integrable multiparametric spin chains \cite{ANG}.


\begin{thebibliography}{99}
\bibitem{CR} I.V. Cherednik, {\em Theor.Math.Phys. 61 (1984) 977 }
\bibitem{SK} E.K. Sklyanin,  {\em J.Phys.A:Math.Gen. 21 (1988) 2375 }
\bibitem{NE} L. Mezincescu and R.I. Nepomechie, {\em J.Phys.A:Math.Gen.
24 (1991) L17}; 
{\em Int.J.Mod.Phys.A7 (1991) 5231; Int.J.Mod.Phys. A7(1992)5657}
\bibitem{DE} H.J. de Vega and A. Gonzalez-Ruiz, {\em Nucl.Phys.B 417 
(1994) 553;
Mod.Phys.Lett.A 9 (1994) 2207; J.Phys.A 27 (1994) 6129}
\bibitem{MV} C.M. Yung and M.T. Batchelor, {\em Nucl.Phys.B 435 (1995) 430}
\bibitem{QF} A. Fring and R. Koberle, {\em Nucl.Phys.B 419 
(1994) 647}; S. Ghoshal
and A.B. Zamolodchikov, {\em Int.J.Mod.Phys.A 9 (1994) 3841}; R. Sasaki, {\em
hep-th/9311027}
\bibitem{SM} R.E. Behrend, P.A. Pearce and D.L. O'Brien, {\em J.Stat.Phys.84 (1996) 1}; C. Ahn and W.M. Koo, {\em Nucl.Phys.B. 468 (1996) 461}; 
M.T. Bachelor, V. Fridkin, A. Kuniba and Y.K. Zhou, {\em Phys.Lett.B 376 (1996) 266}
\bibitem{EKS} F.H.L. Essler, V.E. Korepin and K. Schoutens, {\em Phys.Rev.Lett.
68 (1992) 2960 }
\bibitem{UMO} A.J. Bracken, M.D. Gould, 
J.R. Links and Y.-Z. Zhang ,{\em Phys.Rev.Lett. 74 (1995) 2768 };
G. Bedurfig and H.Frahm,
{\em J.Phys.A:Math.Gen. 28 (1995) 4453}; R.Z. Bariev, A. Kl\"umper and J. Zittartz, {\em Europhys.Lett. 32
(1995) 85 }; J. Links ,{\em cond-mat/9903274}
\bibitem{MA} M.J.Martins and P.B. Ramos, Phys. Rev. B 561 (1997) 6376
\bibitem{DEG} T. Deguchi, A. Fujii and K. Ito, {\em Phys.Lett.B 238 (1990) 242 }
\bibitem{AU} A.J. Bracken, G.W. Delius, M.D. Gould and Y.Z. Zhang, J.Phys. A: Math. Gen. 27 (1994) 6551;
Z. Maassarani, {\em J.Phys.A:Math.Gen. 28 (1995) 1305 };
M.D. Gould, J.R. Links, Y.-Z. Zhang and I. Tsonhantjis, {\em J.Phys.A:Math.Gen. 30 (1997) 4313}
\bibitem{AU1} A.J. Bracken, X.Y. Ge, Y.Z. Zhang and H.Q. Zhou, {\em Nucl.Phys.B516 (1998) 388};
Y.Z. Zhang and H.Q. Zhou, {\em Phys.Lett.A 244 (1998) 427}; D. Arnaudon, {\em JHEP 12 (1997) 6}
\bibitem{SA} H. Saleur, {\em J.Phys.A:Math.Gen. 32 (1999) L207}
\bibitem{SA1} F. Lesage, H. Saleur, P. Simonetti, {\em Phys.Rev.B37 (1998) 4694}
\bibitem{LI} J.R. Links and M.D. Gould, {\em Int.J.Mod.Phys.B10 (1996) 3461}
\bibitem{AL} F.C. Alcaraz, M.N. Barber, M.T. Batchelor, R.J. Baxter and
G.R.W. Quispel, {\em J.Phys.A:Math.Gen.20 (6397) 1987}
\bibitem{JA} H. Asakawa and M. Suzuki, {\em J.Phys.A:Math.Gen. 29 (1996) 225};
M. Shiroishi and M. Wadati, {\em J.Phys.Soc.Jpn. 66 (1997) 1}
\bibitem{AU2}  H.Q. Zhou, X.Y. Ge, J. R. Links and M.D. Gould, {\em Nucl.Phys.B
546 (1999) 779};
H.Q. Zhou, X.Y. Ge and M. D. Gould, {\em cond-mat/9811049}
\bibitem{CHU} Y. Wang, J.-H. Dai, Z.-N Hu, and F.-C Pu, 
{\em Phys.Rev.Lett. 79 (1997) 1901}; Z.-N. Hu, F.-C. Pu and Y. Wang, {\em
J.Phys. A 31 (1998) 5241}
\bibitem{FRA} H. Frahm and N. Slavnov, {\em cond-mat/9810312}
\bibitem{ANG} A. Foerster, J. Links and I. Roditi, {\em J.Phys.A 31 (1998) 687}
\end{thebibliography}
\end{document}